\RequirePackage{xcolor}
\documentclass[journal,twoside,web]{ieeecolor}
\usepackage{silence}
\usepackage{generic}
\usepackage{amsmath,amssymb,amsfonts}
\usepackage{algorithmic}
\usepackage{graphicx}
\usepackage{algorithm,algorithmic}
\usepackage{hyperref}
\usepackage{textcomp}
\usepackage{pgfplots}
\usepackage{epstopdf}
\pgfplotsset{compat=1.8}
\usepackage[labelformat=simple]{subcaption} 
\usepackage[utf8]{inputenc}
\usepackage{siunitx}
\usepackage{balance}
\def\scalex {1}
\def\scalexx {0.98}

\usepackage[backend=biber, style=ieee, maxnames=35, url=false, doi=true, isbn=false, eprint=false, useeditor=false]{biblatex}
\DeclareBibliographyCategory{dataset} 
\addtocategory{dataset}{das_auditory_2020,fuglsang_eeg_2018}
\AtEveryBibitem{%
\ifcategory{dataset}{%
  \clearlist{language}%
  \clearlist{publisher}%
  \clearlist{editor}%
}{%
\clearlist{language}%
\clearfield{note}%
\clearlist{publisher}%
\clearlist{editor}%
}%
}
\DeclareSourcemap{
  \maps[datatype=bibtex, overwrite]{
    \map{
      \step[fieldset=editor, null]
    }
  }
}

\def\BibTeX{{\rm B\kern-.05em{\sc i\kern-.025em b}\kern-.08em
    T\kern-.1667em\lower.7ex\hbox{E}\kern-.125emX}}
\def\BibTeX{{\rm B\kern-.05em{\sc i\kern-.025em b}\kern-.08em
    T\kern-.1667em\lower.7ex\hbox{E}\kern-.125emX}}
\begin{document}
\title{Scattering Transform for Auditory Attention Decoding}
\author{Ren\'{e} Pallenberg, Fabrice Katzberg, Alfred Mertins, and  Marco Maass
\thanks{Ren\'{e} Pallenberg, Fabrice Katzberg, and Alfred Mertins are with the Institute for Signal Processing, University of Luebeck, 23562 Luebeck, Germany (e-mail: r.pallenberg@uni-luebeck.de; f.katzberg@uni-luebeck.de; alfred.mertins@uni-luebeck.de). Fabrice Katzberg is also with the Department for Vision and Machine Learning, FPI Food Processing Innovation GmbH \& Co. KG, 23556 Luebeck, Germany.}
\thanks{Marco Maass is with the German Research Center for Artificial Intelligence (DFKI), AI for Assistive Health Technologies, 23562 Luebeck, Germany (e-mail: marco.maass@dfki.de). }
}
\newcommand{\CV}{$5\times2$ cv}

\maketitle
\begin{abstract}
The use of hearing aids will increase in the coming years due to demographic change. One open problem that remains to be solved by a new generation of hearing aids is the cocktail party problem. 
A possible solution is electroencephalography-based auditory attention decoding. This has been the subject of several studies in recent years, which have in common that they use the same preprocessing methods in most cases. In this work, in order to achieve an advantage, the use of a scattering transform is proposed as an alternative to these preprocessing methods. The two-layer scattering transform is compared with a regular filterbank, the synchrosqueezing short-time Fourier transform and the common preprocessing. 
To demonstrate the performance, the known and the proposed preprocessing methods are compared for different classification tasks on two widely used datasets, provided by the KU Leuven (KUL) and the Technical University of Denmark (DTU). Both established and new neural-network-based models, CNNs, LSTMs, and recent Transformer/graph-based models are used for classification. 
Various evaluation strategies were compared, with a focus on the task of classifying speakers who are unknown from the training. We show that the two-layer scattering transform can significantly improve the performance for subject-related conditions, especially on the KUL dataset. However, on the DTU dataset, this only applies to some of the models, or when larger amounts of training data are provided, as in 10-fold cross-validation. This suggests that the scattering transform is capable of extracting additional relevant information.
\end{abstract}

\begin{IEEEkeywords}
Auditory Attention Decoding, Scattering Transform, EEG, LSTM
\end{IEEEkeywords}

\section{Introduction}
\label{sec:introduction}

Hearing loss is a widespread health problem affecting an increasing population due to demographic change. Hearing aid users often struggle to understand speech when multiple speakers are present simultaneously—the cocktail party problem \cite{haykin_cocktail_2005}. Auditory attention decoding (AAD) addresses this by identifying the attended speaker through brain-computer interfaces, with electroencephalography-based (EEG) systems showing particular promise for hearing aid integration \cite{fiedler_ear-eeg_2016,kaongoen_future_2023,alickovic_system_2016}.
Current AAD approaches and their limitations:
Early correlation-based methods achieve up to $90\%$ accuracy in two-speaker scenarios but require long decision windows (more than 10 seconds) for reliable performance, limiting real-time applicability \cite{de_cheveigne_decoding_2018}. Recent neural-network approaches improve performance \cite{vandecappelle_eeg-based_2021, accou_modeling_2021, pahuja_xanet_2023, qiu_enhancing_2025,dai_gcanet_2026}, yet most rely on the same preprocessing pipeline, extracting audio envelopes via filterbank decomposition and applying bandpass filtering with artifact removal to EEG \cite{biesmans_auditory-inspired_2016,somers_generic_2018}.

This conventional pipeline has two critical limitations. First, envelope extraction compresses stimuli to low-frequency modulation components, discarding potentially diagnostic time-frequency structure such as hierarchical amplitude modulations and cross-frequency interactions. Second, this representation is particularly vulnerable at short decision windows—precisely where real-time hearing aids require robust performance \cite{de_cheveigne_decoding_2018,hjortkjaer_real-time_2025}. While modern data-driven models no longer necessitate strictly correlation-based pipelines, richer time-frequency features remain underexplored in AAD \cite{qiu_enhancing_2025}.

The Scattering Transform (ST) addresses these limitations through theoretically grounded hierarchical time-frequency decomposition. By cascading complex wavelet convolutions with modulus nonlinearities and low-pass averaging, ST constructs representations that are locally translation-invariant and stable to small temporal deformations \cite{mallat_group_2012}. Unlike plain continuous wavelet transforms—recently shown beneficial for AAD \cite{qiu_enhancing_2025}—ST produces second-order coefficients capturing modulations of modulations, explicitly encoding nested envelope dynamics and finer spectral energy distributions \cite{mallat_group_2012,anden_multiscale_2011}.
These properties align precisely with AAD requirements. Neural speech tracking is nonstationary, spans multiple frequency bands (Delta/Theta for envelope tracking, Alpha/Gamma for modulations), and exhibits inter- and intra-individual temporal variability. ST properties such as local invariance, Lipschitz stability to deformations, and energy conservation address the fragility of correlation-based approaches at short windows \cite{mallat_group_2012,de_cheveigne_decoding_2018,hjortkjaer_real-time_2025}. Practically, ST offers decisive advantages for subject-specific AAD where per-subject data are limited. 

First, the ST imposes strong inductive bias analogous to learned convolutional filters but with mathematically designed kernels—substantially reducing learnable parameters and mitigating overfitting \cite{mallat_group_2012,geirnaert_electroencephalography-based_2021,puffay_relating_2023}. Second, ST demonstrates robust performance in audio domains at short windows under variable conditions \cite{anden_multiscale_2011}, matching real-time AAD latency constraints \cite{hjortkjaer_real-time_2025}. Third, its convolutional structure enables efficient GPU/embedded implementations and seamless integration into deep-learning frameworks.

Models and evaluation: To assess the generalizability and robustness of ST preprocessing across different architectural paradigms, we evaluate it with four established neural network models representing diverse design philosophies in AAD.
First, CNN-C1 is an adaptation of one of the earliest convolutional approaches for AAD, consisting of a shallow CNN extended with a channel-attention mechanism to better weight informative EEG channels \cite{cai_eeg-based_2021}. Second, we employ the dilated convolutional network introduced by Accou et al. \cite{accou_modeling_2021}. Owing to its robustness and large receptive field achieved through exponentially increasing dilation rates, this architecture has been repeatedly used as a strong baseline and served as the winning system in the AAD challenges at ICASSP 2023 and 2024 \cite{thornton_detecting_2024}. Third, we include two recurrent architectures based on Long Short-Term Memory (LSTM) networks: LSTM-2 and LSTM-X. LSTM-based approaches have demonstrated competitive performance in multiple AAD studies and offer computational efficiency compared to larger convolutional models \cite{lu_auditory_2021, pallenberg_lstms_2023}. Finally, to evaluate ST with state-of-the-art methods, we test GCANet \cite{dai_gcanet_2026}, a recently proposed architecture that combines graph convolutional networks for spatial EEG modeling with convolutional cross-attention mechanisms for multimodal fusion.

The evaluation strategy significantly impacts reported AAD performance \cite{geirnaert_electroencephalography-based_2021,puffay_relating_2023, qiu_enhancing_2025, yan_overestimated_2025}. Previous works employ varied data splits (ten random 80/10/10 folds, k-fold cross validation (cv)) with inconsistent significance testing, complicating cross-study comparisons \cite{puffay_relating_2023}. We therefore adopt Dietterich's \CV{}, which provides high statistical reliability through paired testing across multiple data splits, with computational effort comparable to or less than ten random folds \cite{dietterich_approximate_1998}. We additionally compare \CV{} against k-fold cv with varying $k \in \{3,4,5,10\}$ and speaker-wise validation to systematically analyze how evaluation protocols influence performance assessment and model selection.

This work makes three primary contributions:
\begin{enumerate}
\item We systematically investigate two-layer ST as preprocessing for AAD, comparing it against three alternative approaches: the conventional baseline (gammatone filterbank with envelope extraction), the synchrosqueezed short-time Fourier transform (SSQ-STFT), and regular filterbank preprocessing. We demonstrate ST's effectiveness across multiple neural architectures on two public datasets (KUL, DTU).
\item We conduct comprehensive evaluation strategy comparisons (Dietterich's \CV{}, k-fold cv with $k \in \{3,4,5,10\}$, speaker-wise validation) with rigorous significance testing to identify protocols most appropriate for AAD scenarios.
\item We provide detailed computational analysis of ST (runtime, number of floating point operations (FLOPs), implementation considerations) for practical deployment assessment.
\end{enumerate}
\section{Methods}
\subsection{Scattering Transform}
Auditory attention operates across multiple temporal scales: the brain tracks fast phonemic features ($\approx$20-50 ms), syllabic rhythms ($\approx$200 ms), and prosodic patterns ($>$500 ms) simultaneously. Traditional filterbanks capture spectral content but collapse temporal modulations within each band. The ST \cite{mallat_group_2012} addresses this limitation by computing multi-order modulation representations through cascaded wavelet filtering.

The ST applies two sequential filterbanks of complex Morlet wavelets with logarithmically-spaced center frequencies, separated by a modulus operation. For input signal $x(t)$, the two scattering layers are:
\begin{align}
S_1(x,\lambda_1) &= |x\star\psi_{\lambda_1}|\star\phi, \\
S_2(x,\lambda_1,\lambda_2) &= ||x\star\psi_{\lambda_1}|\star\psi_{\lambda_2}|\star\phi,
\end{align}
where $\psi_{\lambda_\ell}$ are wavelets at center frequencies $\lambda_\ell$, $\phi$ is a low-pass averaging filter, and $\star$ denotes convolution. Additionally, the low-pass-filtered input $S_0(x)=x \star \phi$ is retained. The number of filters per octave $Q$ and the scale parameter $J$ that determines the downsampling factor $2^J$ are varied for different experiments. The Python library by Andreux et al. was used as implementation \cite{andreux_kymatio_2020}. 

The first layer provides a standard wavelet filter bank decomposition, a form of time-frequency representation. Additionally, the second layer captures modulations of these modulations—second-order temporal structure that characterizes, for example, how the speech envelope itself varies. This hierarchical representation aligns with the multiscale nature of cortical speech tracking and provides information inaccessible to conventional first-order filterbank approaches. 
\begin{figure}
\includegraphics[scale=1]{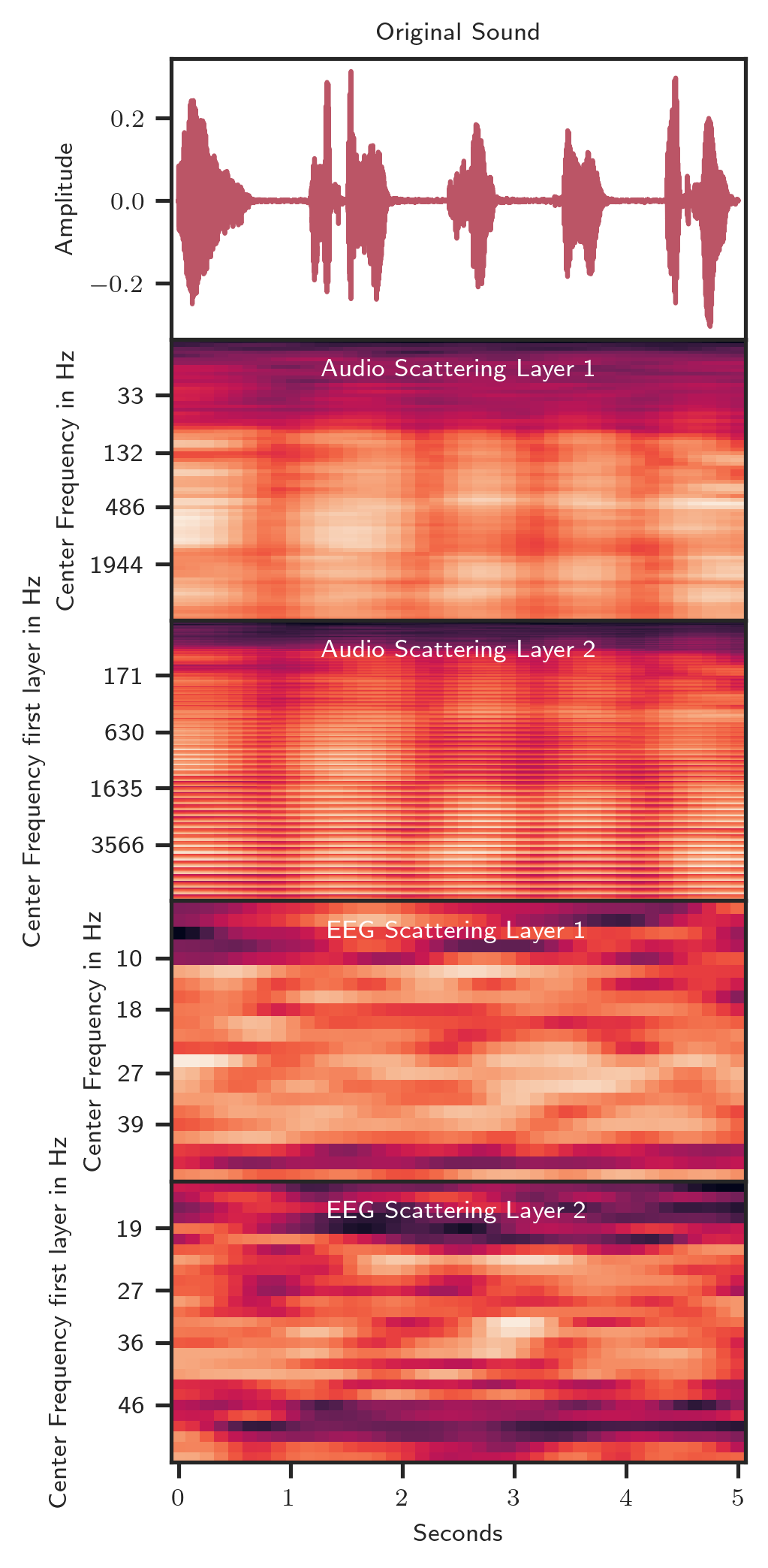}
  \caption{Scattering transform applied to audio (top) and EEG at T8 (bottom) from KUL dataset. Layer 2 captures fine-grained temporal modulations invisible in Layer 1, shown by temporal striations (0.0-0.5s) and differentiated patterns during acoustically similar events (3.5-4.5s). Original signal and ST layers 1-2 with, $F_o=8$, $Q=8$.}
  \label{fig:scat_example}
\end{figure}

To demonstrate the second layer's additional information content, in \autoref{fig:scat_example}, we examine two key regions for both audio and EEG signals. First, the initial transient (0.0-0.5s) appears as a coarse, smeared event in Layer 1 for both modalities, while Layer 2 resolves fine temporal striations revealing individual modulation cycles invisible in Layer 1 (particularly visible in audio coefficients 630-1635 Hz and EEG coefficients 27-36 Hz). Second, two events at 3.7s and 4.5s appear similar in Layer 1 but exhibit distinctly different activation patterns in Layer 2, with the former showing pronounced activity in higher coefficients (audio: 1635-3566 Hz). Additionally, Layer 2 captures transition dynamics around 4.0s barely perceptible in Layer 1. These observations demonstrate that Layer 2 provides both superior temporal resolution and second-order modulation information for both audio and neural signals—capturing the nested temporal structure essential for discriminating acoustically similar events and their corresponding neural responses.

\subsection{Synchrosqueezing Short-Time-Fourier Transform}
The Synchrosqueezing short-time-Fourier transform (SSQ-STFT) is another advanced method to get a time-frequency representation of a signal \cite{oberlin_fourier-based_2014}. It enhances the classical STFT by reducing spectral smearing and improving concentration of energy around instantaneous frequencies. The SSQ-STFT first computes the standard STFT and then estimates the local instantaneous frequency from the phase evolution of each time-frequency coefficient. Using this estimate, the algorithm reassigns the STFT energy along the frequency axis (“synchrosqueezing”), yielding a sharper and more interpretable representation that remains mathematically invertible.
This method is particularly advantageous for audio and EEG signals, both of which are highly nonstationary and contain multiple overlapping oscillatory components.


\subsection{Models}
\begin{figure}
\begin{subfigure}{0.49\textwidth}
\scalebox{0.85}{\includegraphics{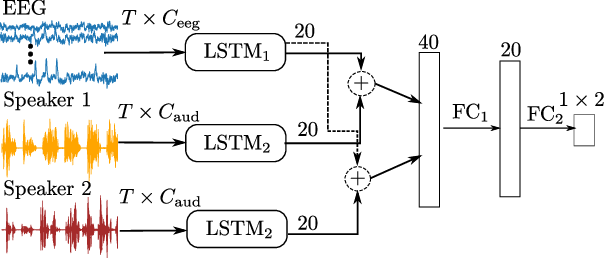}}
\caption{LSTM-2}
\label{fig_lstm}
\end{subfigure}
\begin{subfigure}{0.5\textwidth}
\scalebox{0.85}{\includegraphics{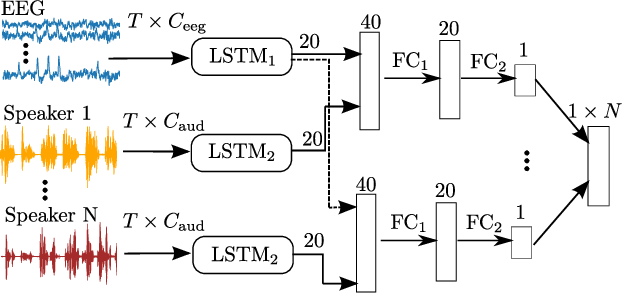}}
\caption{LSTM-X}
\label{fig_multilstm}
\end{subfigure}
\caption{Architecture of the models used. The input has $T$ time steps, $C_\mathrm{eeg}$ channels for the EEG, and $C_\mathrm{aud}$ channels for the audio signals. The inputs are on the left, and the output is on the right. LSTM-2 can only work with two speakers, while the LSTM-X can handle a variable number of speakers.}
\label{fig_models}

\end{figure}

\label{met_models}
\subsubsection{CNN-C1}
\begin{table}
\centering
\caption{CNN-C1: Detailed layer configuration. AvgTime eliminates the time dimension by averaging over $T$. Convolutions use no padding and stride $1$.}
\begin{tabular}{|l|c|c|}
  \hline

  \multicolumn{3}{|c|}{EEG $x_e \in \mathbb{R}^{T \times 64C_e}, \text{Audio}_1 x_1, \text{Audio}_2 x_2 \in \mathbb{R}^{T \times C_a}$}\\
\hline
Input & Steps & Output \\
\hline

\hline
 $x_e$ & AvgTime 
  & $w_e \in \mathbb{R}^{1 \times 64C_e}$ \\

$w_e$ 
  & Dense(units=8, tanh), DO(0.5) 
  & $w_e \in \mathbb{R}^{8}$ \\

$w_e$ 
  & Dense(units=64$C_e$, tanh) 
  & $w_e \in \mathbb{R}^{64C_e}$ \\
\hline

$x_e$
  & $x_e\odot w_e$, 
  & $x_e \in  \mathbb{R}^{T \times 64C_e} $ \\
\hline

  & Concat($x_e,x_1,x_2$) 
  & $x \in \mathbb{R}^{T\times(64C_e+2C_a)}$ \\
\hline

$x$ 
  & Conv1D(kernel=9, filters=10, ReLU) 
  & $x \in \mathbb{R}^{(T-8)\times 10}$ \\

$x$ 
  & AvgTime 
  & $x \in \mathbb{R}^{10}$ \\

$x$ 
  & Dense(units=10, ReLU), DO(0.5) 
  & $x \in \mathbb{R}^{10}$ \\

$x$ 
  & Dense(units=2, Sigmoid) 
  & $Y \in [0,1]^2$ \\
\hline
\end{tabular}
\label{tab:cnn_c1_conf}
\end{table}
The first baseline model is based on the work of Cai et al. \cite{cai_eeg-based_2021} and includes a CNN developed by Van de Cappelle et al. \cite{vandecappelle_eeg-based_2021}. This model is enhanced with a channel attention mechanism (C1) applied to the EEG channels. For more details, refer to the original publication \cite{cai_eeg-based_2021} and \autoref{tab:cnn_c1_conf}.

\subsubsection{CNN-Dil}
\begin{table}
\centering
\caption{CNN-Dil: Detailed layer configuration. Strides are always 1. Dilation (dil) defaults to 1 unless specified. For signals with $T<16$, dil=1 in all convolutions. Kernel length (k) is at the first postion, number of filters (f) at the second. }
\begin{tabular}{|l|c|c|}
   \hline

  \multicolumn{3}{|c|}{EEG $x_e \in \mathbb{R}^{T \times 64C_e}, \text{Audio}_i x_i \in \mathbb{R}^{T \times C_a}$}\\
\hline
\hline
Input & Steps & Output \\
\hline

$x_e$  & Conv1D(k=1, f=8) 
  & $x_e \in \mathbb{R}^{T \times 8}$ \\

$x_e$ 
  & Conv1D(k=3, f=16, ReLU) 
  & $x_e \in \mathbb{R}^{(T-2)\times 16}$ \\

$x_e$ 
  & Conv1D(k=3, f=16, ReLU, dil=3) 
  & $x_e \in \mathbb{R}^{(T-8)\times 16}$ \\

$x_e$ 
  & Conv1D(k=3, f=16, ReLU, dil=9) 
  & $x_e \in \mathbb{R}^{(T-15)\times 16}$ \\
\hline

$x_i$ 
  & BN, Conv1D(k=3, f=16, ReLU) 
  & $x_i \in \mathbb{R}^{(T-2)\times 16}$ \\

$x_i$ 
  & Conv1D(k=3, f=16, ReLU, dil=3) 
  & $x_i \in \mathbb{R}^{(T-8)\times 16}$ \\

$x_i$ 
  & Conv1D(k=3, f=16, ReLU, dil=9) 
  & $x_i \in \mathbb{R}^{(T-15)\times 16}$ \\
\hline

$(x_e,x_i)$ 
  & CosineSimilarity$(x_e,x_i)$ 
  & $x_i \in \mathbb{R}^{16\times 16}$ \\
\hline

$x_i$ 
  & Flatten, Dense(units=1) 
  & $y_i \in [0,1]$ \\

$[y_i,\dots,y_n]$ 
  & Softmax($[y_1,\dots,y_n]$) 
  & $Y \in [0,1]^n$ \\
\hline
\end{tabular}
\label{tab:cnn_dil_conf}
\end{table}
The CNN-Dil model, developed by Accou et al. \cite{accou_modeling_2021}, features multiple pathways where an individual output is predicted for each audio signal, independent of other audio signals. For more details, refer to the original publication \cite{accou_modeling_2021} and \autoref{tab:cnn_dil_conf}.

\subsubsection{LSTM-2}
\begin{table}
\centering
\caption{LSTM-2: Detailed layer configuration}
\begin{tabular}{|l|c|c|}
\hline
Input & Steps & Output \\
\hline
EEG $\in \mathbb{R}^{T\times 64C_e}$ 
  & BN, LSTM$_1$(units=15) 
  & $x_e \in \mathbb{R}^{15}$ \\

Audio$_1 \in \mathbb{R}^{T\times C_a}$ 
  & BN, LSTM$_2$(units=15) 
  & $x_1 \in \mathbb{R}^{15}$ \\

Audio$_2 \in \mathbb{R}^{T\times C_a}$ 
  & BN, LSTM$_2$(units=15) 
  & $x_2 \in \mathbb{R}^{15}$ \\
\hline

$(x_e,x_1, x_2)$ 
  & Concat($x_e+x_1,\; x_e+x_2$) 
  & $x \in \mathbb{R}^{30}$ \\
\hline

$x$ 
  & BN, DO(0.5)& $x \in \mathbb{R}^{30}$ \\
  $x$ & Dense (units=40, ReLU)  & $x \in \mathbb{R}^{40}$ \\

$x$ 
  & BN, DO(0.5)
  & $x \in \mathbb{R}^{40}$ \\

$x$ 
  & Dense (units=2, Sigmoid)
  & $Y \in [0,1]^{2}$ \\
\hline
\end{tabular}
\label{tab:tlstm_2_conf}
\end{table}
The LSTM-2 architecture, inspired by Lu et al. \cite{lu_auditory_2021}, uses two distinct LSTM cells (15 hidden units each): one processes the EEG signal, the other processes each audio signal individually. The EEG output is combined with each audio output (yielding two 15-dimensional vectors), concatenated, and fed through two fully connected layers (40 units with ReLU, two units with sigmoid function). Batch normalization precedes each LSTM cell and dense layer; $50\%$ dropout (DO) is applied before dense layers to prevent overfitting. See \autoref{fig_lstm} for architecture visualization and detailed layer configuration in \autoref{tab:tlstm_2_conf}.
 
\subsubsection{LSTM-X}
\begin{table}
\centering
\caption{LSTM-X: Detailed layer configuration.}
\begin{tabular}{|l|c|c|}
\hline
Input & Steps & Output \\
\hline
EEG $\in \mathbb{R}^{T\times64C_e}$  & BN, LSTM$_1$(units=15) &$x_e \in \mathbb{R}^{15}$ \\
Audio$_i\in \mathbb{R}^{T\times C_a}$  & BN, LSTM$_2$(units=15) & $x_i \in \mathbb{R}^{15}$ \\
\hline
$x_e,x_i$ & Concat($x_e,x_i$), BN, DO(0.5) & $x_i \in \mathbb{R}^{30} $\\
 \hline
$x_i$ & Dense(units=20, ReLU)& $x_i \in \mathbb{R}^{20} $ \\
$x_i$ &BN, DO(0.5) & $x_i \in \mathbb{R}^{20} $ \\
$x_i$ & Dense(units=1, Sigmoid) & $y_i \in [0,1] $\\
$[y_i,...,y_n]$ & Softmax($[y_1,\dots,y_n]$) &$Y \in [0,1]^{n} $\\
\hline
\end{tabular}
\label{tab:lstm_x_conf}
\end{table}
LSTM-X extends LSTM-2 to handle  $n$ audio signals through a multipath architecture. Two LSTM cells (15 hidden units each) reduce the time dimension of EEG and each audio signal. For each audio signal, its output is concatenated with the EEG output and processed by two fully connected layers (20 units with ReLU, one unit with sigmoid). Batch normalization and DO are applied identically to LSTM-2. Individual outputs are merged via softmax to produce final predictions (see \autoref{fig_multilstm}). Unlike LSTM-2, LSTM-X and CNN-Dil can handle more than two audio signals without architectural modifications or retraining. Detailed layer configurations are provided in \autoref{tab:lstm_x_conf}.

\subsubsection{GCANet}
GCANet integrates EEG and audio through graph-based EEG modeling, convolutional audio encoding, and crossmodal attention \cite{dai_gcanet_2026}. The EEG encoder processes raw EEG plus 10 engineered features per channel, namely differential entropy and power spectral density for five frequency bands ([4-8], [8-12], [12-30], [30-50] Hz). Relative spatial positions of EEG channels are encoded to guide graph construction. A Graph Convolutional Network (GCN) produces a $64 \times 128$ representation. 

The audio encoder takes 16 kHz speech envelopes, downsamples to 128 Hz through stacked convolutional layers, and expands to 64 feature channels, yielding a $64 \times 128$ representation. 

Encoded modalities are fused using four stacked convolutional cross-attention modules that link EEG (conditioning target) with each audio signal (context) independently. The final EEG output is concatenated with the initial EEG features along the channel dimension, passed through 1D convolution to obtain the multimodal feature map and classified via two fully connected layers. Further architectural details are available in \cite{dai_gcanet_2026}.

\subsubsection{GCANet-NoEn}
\begin{table}
\centering
\caption{EEGEncoder: GCANet-NoEn.}
\begin{tabular}{|l|c|c|}
     \hline
    \multicolumn{3}{|c|}{EEG $x_e \in \mathbb{R}^{T \times 64 \times C_e}$}\\
\hline
\hline
Input & Steps & Output \\
\hline

-  & $A = A_{ds} + A_{learned}$ 
  & $A \in \mathbb{R}^{64 \times 64}$ \\

-  & Normalized Laplacian $L = AD^{-1}$ 
  & $L \in \mathbb{R}^{64 \times 64}$ \\
\hline

$X_e$ 
  & ELU$(X_eL + X_e)$ 
  & $x_e \in \mathbb{R}^{T \times 64 \times C_e}$ \\
\hline

$x_e$ 
  & Reshape 
  & $x_e \in \mathbb{R}^{64 \times TC_e}$ \\
\hline

$x_e$ 
  & Dense(units=128), DO(0.5) 
  & $x_e \in \mathbb{R}^{64 \times 128}$ \\
\hline
\end{tabular}
\label{tab:eeg_encoder}
\end{table}

\begin{table}
\centering
\caption{AudioEncoder: GCANet-NoEn.}
\begin{tabular}{|l|c|c|}
     \hline
  \multicolumn{3}{|c|}{$\text{Audio}_i x_i \in \mathbb{R}^{T \times C_a}$}\\
\hline
\hline
Input & Steps & Output \\
\hline
$x_i$ 
  & Conv1D(filters=64, kernel=8), BN, ReLU 
  & $x_i \in \mathbb{R}^{64 \times C_a}$ \\

$x_i$ 
  & Dense(units=128), BN, ReLU 
  & $x_i \in \mathbb{R}^{64 \times 128}$ \\
\hline
\end{tabular}
\label{tab:audio_encoder}
\end{table}
The original GCANet contains encoding blocks for audio and EEG signals. For ST and SSQ-STFT preprocessed signals, these encoding blocks were removed and replaced with conversion blocks to adapt signals to match the classifiers input dimensions of $64 \times 128$.

For EEG $x_e \in \mathbb{R}^{T \times 64 \times C_e}$, with 64 EEG channels, $T$ time samples and $C_e$ feature channels, the position graph encoding from the original GCANet is retained, which encodes relative EEG channel positions through matrix multiplication. Time and channel dimensions are then combined and projected through a linear layer (see \autoref{tab:eeg_encoder}).
For audio signals $x_i \in \mathbb{R}^{T \times C_a}$, with $C_a$ feature channels, a 1D convolution layer (64 filters, kernel size 8) convolves across the channel dimension, expanding the time dimension to 64. This is followed by a linear layer, mapping the number of channels to 128 (see \autoref{tab:audio_encoder}).
These adapted signals are fed into the unchanged convolutional cross-attention mechanism of the original GCANet \cite{dai_gcanet_2026}.
\section{Evaluation}

\subsection{KU Leuven Dataset}

The Department of Neuroscience at KU Leuven collected a dataset comprising data from 16 individuals with normal hearing abilities \cite{das_effect_2016, das_auditory_2020}. The study utilized audio stimuli consisting of four audiobooks. From each book, two six-minute segments were taken. Each participant was randomly assigned to two of these stories, ensuring exposure to only two distinct speakers simultaneously. Initially, participants listened to two segments of one story, with alternating conditions and stimulated ears. The same stories were replayed, but this time, participants were instructed to focus on the other story. This procedure was repeated with the remaining two stories, resulting in eight unique presentations totaling 48 minutes. For this work each of the eight trials was split into seven trials of 52 seconds length to nearly fit the trial length of the DTU dataset.  
 EEG data was recorded using a 64-channel EEG system with a sampling rate of 8192 Hz. The data was pre-processed, whereby the EEG signals were filtered with a high-pass filter with a cut-off frequency of 0.5 Hz and then downsampled to 128 Hz. Additionally, artifacts were removed with a Wiener filter \cite{somers_generic_2018}.

\subsection{DTU Dataset}
The DTU compiled a dataset that includes recordings from 18 individuals with normal hearing abilities \cite{fuglsang_eeg_2018}. The participants were exposed to two naturally spoken stories, each divided into 50-second segments. Each participant underwent 60 trials where two audio stimuli were played simultaneously, and they were instructed to concentrate on one of the stories. EEG signals were recorded from the participants during these trials. There was a pause between each segment, and the directions of the stimuli were randomly altered. 
A 64-channel EEG system with a sampling rate of 512 Hz was used to record the EEG data. In parallel, six EOC channels were recorded to track eye movement, enabling artifact reduction \cite{wong_comparison_2018}.
\subsection{Preprocessing}
\subsubsection{Baseline}
\label{prep_baseline}
Pipeline 1 adopted the standard preprocessing techniques used in multiple works \cite{vandecappelle_eeg-based_2021,cai_eeg-based_2021}. Initially, the EEG channels were re-referenced to the mean of all channels. The eye artifacts were then reduced. The method used depends on the dataset. The DTU dataset contains EOC channels for tracking eye movements. These can be used to reduce the artifacts \cite{de_cheveigne_sparse_2016}.
The KUL dataset lacks these additional channels. Therefore, a generalized Wiener filter was applied \cite{somers_generic_2018}. Subsequently, the signals were bandpass filtered between 1 Hz and 32 Hz and downsampled to a rate of 64 Hz. All filters were applied in the frequency domain without phase shift.

For the audio stimuli, the auditory-inspired method developed by Briesman et al. was employed \cite{biesmans_auditory-inspired_2016}. The rationale behind this filter bank is to mimic the processing of the human auditory system. The audio signal was decomposed into 28 sub-bands using a gammatone filter bank, with center frequencies uniformly distributed on the ERB scale between 50 Hz and 5000 Hz. In this work, the filters were implemented using the Python SciPy library and the COCOHA Matlab Toolbox \cite{wong_comparison_2018}. These sub-band envelopes were then raised to the power of 0.6 and subsequently summed. Finally, the result was filtered between 1 Hz and 32 Hz using a Butterworth bandpass filter and downsampled to 64 Hz.
\subsubsection{Scattering-pipeline}
\label{prep_scat}
As a first step, the audio signals were resampled from 44100 Hz to 16384 Hz in order to obtain a power of two. The same procedure was applied to the EEG signals, which were downsampled to 128 Hz.
The implementation applies the filter banks of the ST in the frequency domain \cite{andreux_kymatio_2020}. To keep the processing delay moderate, the EEG and audio signals were divided into one-second segments, as the implementation does not provide an efficient implementation for long windows such as overlap-add. These segments were then individually processed by the ST. Finally, the results were concatenated. The number of filters for EEG and audio processing, given by $Q_e$ and $Q_a$, respectively,  as well as the output frequency ${F}_{o}$ were varied for the different experiments. Note that the number of output channels is defined by these parameters, see Tables \ref{Tab:qfilter_audo} and \ref{Tab:qfilter_eeg}.
\begin{table}
\centering
\caption{Number of channels after the ST for one audio channel with 16384 Hz sampling rate for different $F_o$, $Q$ and $J$. The number of channels in each of the three scattering layers is listed separately.}
\begin{tabular}{c|c|c||c|c|c|c|c|c}
$F_o$ &$J$ & $Q$ & 1 & 2 & 3 & All & FLOPs& Lag\\
\hline
64 & 6 & 8 & 1 & 54 & 179 & 234 & 11M & 0.06 \\
32 & 7 & 8 & 1 & 62 & 237 & 300 & 12M & 0.13\\
16 & 8 & 8 & 1 & 70 & 303 & 374 & 15M & 0.25\\
8 &9 & 8 & 1 & 78 & 377 & 456 & 20M & 0.5\\
\hline
8 & 11& 4 & 1 & 42 & 189 & 232  & 10M& 0.5\\
8 & 11& 2 & 1 & 23 & 108 & 132  & 7M& 0.5\\
8 & 11& 1 & 1 & 12 & 63 & 76  & 5M& 0.5 \\
\end{tabular}
\label{Tab:qfilter_audo}
\end{table}
\begin{table}
\centering
\caption{Number of channels after the ST for one EEG channel with 128 Hz sampling rate for different $F_o$, $Q$ and $J$. The number of channels in each of the three scattering layers is listed separately.}
\begin{tabular}{c|c|c||c|c|c|c|c|c}
$F_o$ & $J$& $Q$ & 1 & 2 & 3 & All & FLOPs & Lag\\
\hline
64 &1& 8 & 1 & 7 & 0 & 8 & 16,173 & 0.06\\
32 &2& 8 & 1 & 7 & 0 & 8 & 8,354 & 0.13\\
16 &3& 8 & 1 & 14 & 9 & 24 & 36,734 & 0.25 \\
8 & 4& 8 &1 & 22 & 27 & 50 & 69,724 & 0.5 \\
\hline
8 &  4 & 4 & 1 & 14 & 14 & 29 & 38,219& 0.5  \\
8 &  4 & 2 & 1 & 8 & 11 & 20 &27,766& 0.5 \\
8 &  4 & 1 & 1 & 5 & 7 & 13& 19,602& 0.5 \\
\end{tabular}
\label{Tab:qfilter_eeg}
\end{table}
\subsubsection{Synchrosqueezing short-time Fourier transform}
To contextualize the ST's performance, we compare against the SSQ-STFT \cite{oberlin_fourier-based_2014}. The SSQ-STFT was configured to match the ST's temporal resolution: hop sizes of 512 samples for audio \mbox{(32 Hz at 16384 Hz)} and 16 samples for EEG  \mbox{(8 Hz at 128 Hz)}, with window sizes of 1024 and 64 samples respectively. The implementation from Muradeli et al. was used \cite{muradeli_overlordgolddragonssqueezepy_2024}. Unlike the ST's multi-order modulation structure, SSQ-STFT provides enhanced frequency localization within a single-layer time-frequency decomposition. This configuration is not applicable to the CNN-C1 and CNN-dil because they need identical sampling frequencies for the audio and EEG. 
\subsection{Data Splitting}
\label{sec_split}
Common evaluation strategies for machine learning in auditory attention decoding have been reviewed \cite{puffay_relating_2023, qiu_enhancing_2025}.  It has been demonstrated that the results of a model is strongly influenced by the strategy employed for splitting the training and test data. Various paradigms for this will be described in detail in the sequel.
\subsubsection{Full-shuffle}
In the full-shuffle mode, a model is trained and validated independently for each subject. All trials from the experiments are combined. From this data, random windows are selected as validation data, while the remainder is used as training data. This method has been frequently employed in previous works that apply neural networks in AAD \cite{cai_eeg-based_2021, lu_auditory_2021, cai_eeg-based_2024}. However, it has been shown that this method does not generalize well, and alternative methods should be considered \cite{puffay_relating_2023}.
\subsubsection{Trial-wise}
In the trial-wise mode, each subject has a dedicated model trained individually. The trials from the experiments are not split up, but assigned as a whole to either the training or the validation dataset. By preventing the models to learn from neighboring windows, this method reduces the risk of overfitting to local patterns. It allows for a more reliable assessment of the model's generalization capabilities \cite{puffay_relating_2023}.
\subsubsection{Speaker-wise}
\label{sec:speaker-wise}
The complexity of the model's task increases if the speakers used for training are not included in the test dataset. The speaker-wise mode is similar to the trial-wise paradigm, but with one crucial difference: all data of a single speaker is either in the training or in the validation dataset. This approach is only applicable to the KUL dataset, as the DTU dataset only contains two speakers. In the KUL dataset, half of the trials are performed by two other competing speakers than the other half. Therefore, only two splits can be evaluated in this scenario. This method ensures a more robust evaluation of the model's ability to generalize across different speakers.
\subsubsection{Cross-subject}
When evaluating across different subjects, a model is trained with the data of several subjects and then validated on one or multiple other subjects. The major advantage of a cross-subject model is that it eliminates the need for individual training for each subject. This would be of great advantage in real-life applications, as it allows for a scalable and efficient use of the model.
\subsubsection{Cross-subject and speaker-wise}
The most generalizing approach combines the cross-subject and speaker-based paradigms, in which the model is evaluated on both an unseen subject and an unseen speaker. However, this method reduces the number of training examples that can be used for a single model by half compared to the simple cross-subject mode. This approach provides the most stringent test of the model's generalization ability, but also poses a greater challenge due to the smaller amount of training data. Nevertheless, it could lead to a model that is very versatile and can be used in a variety of real-world scenarios.
\subsection{Validation}
\label{sec:val}
The goal of this work is to compare the performance of different preprocessing-pipelines. Therefore, validation methods were applied which allow for significance tests of high statistical power. 
There are a variety of validation strategies that have been used in previous works, including five-fold cv, ten-fold 80/20 splits, etc. \cite{puffay_relating_2023}. No gold standard has yet been developed in the field of AAD, which makes comparison with previous works difficult. 
 
\subsubsection{Dietterich's cv}
One validation method with a high statistical power is Dietterich's \CV{} \cite{dietterich_approximate_1998}. To compare two classifiers, the data is divided into two subsets. One is used for training and the other for validation. The difference between the two validation results is stored. As with k-fold cv, the subsets swap places for a second run. This is performed five times with different random splits. Thereby ten difference scores $p_i^{(j)}$ are generated for the runs $i$ and splits $j$. The mean of one run is given by $\bar{p_i}=\frac{p_i^{(1)}+p_i^{(2)}}{2}$. With these values a paired t-test can be performed, where the $t$-value is defined as:
\begin{equation}
t= \frac{p_1^{(1)}}{\sqrt{\frac{1}{5}\sum_{i=1}^5 (p_i^{(1)}-\bar{p_i})^2+(p_i^{(2)}-\bar{p_i})^2}}.
\end{equation}

\subsubsection{K-fold cross-validation}
For a k-fold cv, $t$ is given by: 
\begin{equation}
t= \frac{\bar{p}\times\sqrt{n}}{\sqrt{\frac{1}{n-1}\sum_{i=1}^{n-1} (p^{(i)}-\bar{p})^2}},
\end{equation}
where $\bar{p}$ is the mean score over all splits and $p^{(i)}$ is the score for split $i$. 
It had been shown that the probability of errors is lower with \CV{} than with k-fold cv with ten folds. In the context of large deep neural networks, evaluation time is an issue. The \CV{} has the same number of runs but with less data, which leads to a reduction in training-time by $\frac{k-1}{5}$. 

The \CV{} was used in this work for the division of the trials in the trial-wise evaluation as well as for the division of the subjects for the cross-subject tests. This method could not be used for the speaker-wise division, so the significance test for k-fold cv was used. 
If significance is claimed for the results, the tests described here were applied with a significance level of $\alpha=0.01$.

\subsection{Training Parameters}
All models except GCANet and GCANet-NoEn were implemented in Keras/TensorFlow 2.11.10 (Python 3.10.11) and trained for up to 150 epochs with early stopping (patience: 5 epochs, minimum: 10 epochs). Training data was augmented using a sliding window approach. The Adam optimizer was used with an initial learning rate of 0.001, reduced by 0.5 on plateau (patience: 3 epochs).
GCANet and GCANet-NoEn were implemented in PyTorch 2.10 following the original configuration \cite{dai_gcanet_2026}: 30 epochs, AdamW optimizer (weight decay: 0.0001), initial learning rate of 0.0001 reduced by 0.8 on plateau (patience: 2 epochs).
All models used categorical cross-entropy as loss function.

\section{Results and Discussion}
The pipelines and models were tested with different configurations, which differ in terms of the dataset and the evaluation method.

\subsection{Trial-Wise Experiments}
For the trial-wise validation the trials were split into training and test following the \CV{} method resulting in ten training and validation runs per subject. The first set of experiments was conducted subject-wise and not across subjects. Individual classifiers were trained and tested for each subject.

The four models described were evaluated for four different window lengths ($L_x$) of 0.25, 0.5, 1 and 2 seconds (\si{\second}), as far as this was possible. The CNN-Dil and CNN-C1 models require a minimum number of samples, which was not achieved for all tests. Both the baseline and scattering  pipelines were tested for all these configurations.
\subsubsection{Baseline}
In most previous works with the baseline preprocessing, a full shuffle evaluation was done \cite{cai_eeg-based_2021,lu_auditory_2021,cai_eeg-based_2024}. The presented accuracies of up to 90\% could not be repeated with the trial-wise evaluation on the same datasets. This has already been shown in previous works and underlines the need for more robust evaluation methods in the field of AAD \cite{vandecappelle_eeg-based_2021, puffay_relating_2023, qiu_enhancing_2025}. 

On both datasets, no median accuracies above 0.64 were achieved with the baseline pipeline. Especially in \cite{qiu_enhancing_2025} it is shown that on the DTU dataset with a trial-wise evaluation only results were achieved that are marginally better than chance. This pattern was reproduced in this work, see \autoref{fig:ModelsDTUTrialwise}. 

\subsubsection{Scattering transform and SSQ-STFT}

\begin{figure}
\begin{subfigure}{0.49\textwidth}
\scalebox{1}{ \includegraphics[scale=1]{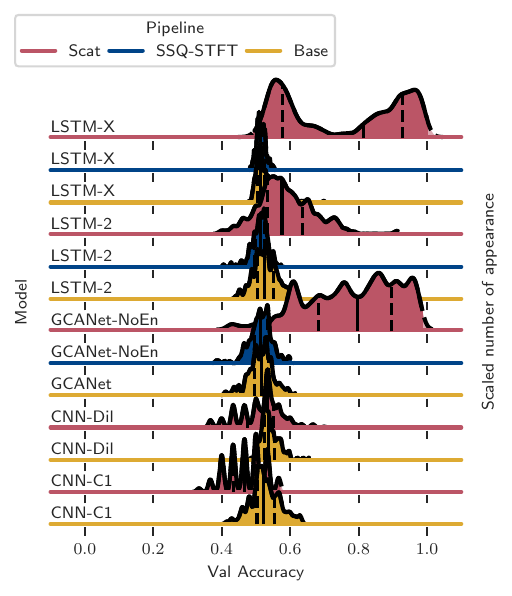} 
}
\caption{DTU}
\label{fig:ModelsDTUTrialwise}
\end{subfigure}
\begin{subfigure}{0.49\textwidth}
\scalebox{1}{ \includegraphics[scale=1]{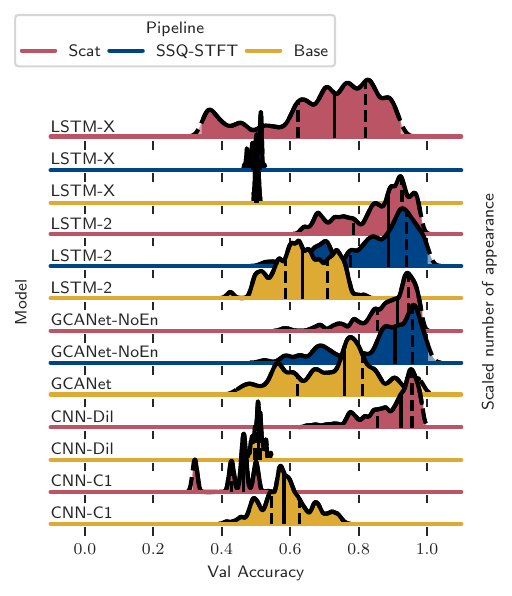} 
}
\caption{KUL}
\label{fig:ModelsKULTrialwise}
\end{subfigure}
\caption{Kernel density plots comparing baseline (base), ST (scat), and SSQ-STFT preprocessing on DTU and KUL datasets. Accuracy distributions for trial-wise \CV{} across all subjects with $L_x=\SI{2}{\second}$
(CNN-C1, CNN-Dil) and $L_x=\SI{1}{\second}$ (others). ST parameters: $F_o=\SI{8}{\Hz}$, $Q_a=Q_e=8$.
}
\label{fig:ModelsTrialwise}
\end{figure}

The ST was tested with different configurations for $F_o$ and a constant value of eight for $Q_a$ and $Q_e$. The results for $F_o$ of 8 Hz and 16 Hz are the best and almost identical. The results with 8 Hz are therefore discussed in particular below. In \autoref{fig:ModelsTrialwise}, the results for the baseline preprocessing, the SSQ-STFT, and the ST are compared.

The ST and the SSQ-STFT improve performance on the KUL dataset across nearly all models, achieving mean accuracies of up to 0.92. However, on the DTU dataset with \CV{} evaluation, only GCANet-NoEn and LSTM-X show substantial improvements. For SSQ-STFT and the baseline pipeline, all models achieve only chance-level performance.

These findings align with literature reports of worse DTU performance compared to KUL, often at chance level. Qiu et al. achieved maximum mean accuracy of 0.55 using 4-fold cv \cite{qiu_enhancing_2025}, consistent with our maximum of 0.60. To investigate the influence of training data quantity, k-fold cv was tested with $k \in \{3,4,5,10\}$. Results reveal a clear data-efficiency trend: accuracy increases monotonically with training set size, reaching $\approx$ 0.84 at 5-fold cv with LSTM-2, see \autoref{Tab:resultseavaluation}. This indicates that training data availability, rather than inherent ST limitations, is the primary constraint.

Two dataset-specific factors explain the DTU-KUL discrepancy. First, we look at the density of attention changes. The KUL data contains 8 switches per recording, yielding longer stationary segments, while DTU data contains 60 attention changes, producing shorter stable contexts and more boundary regions where performance degrades, i.e., accuracy empirically drops at segment beginnings and ends. A second factor is the acoustic heterogeneity. DTU data spans three distinct acoustic environments, introducing domain shifts that reduce feature effectiveness under quasi-stationary assumptions.
Despite these challenges, the ST achieves strong results with multiple models on DTU, distinguishing it from SSQ-STFT and the baseline pipeline.
\subsection{Speaker-Wise Evaluation on KUL Dataset}
For the real-world application, the speaker-wise evaluation provides the more interesting results, see Section \ref{sec:speaker-wise}. 
Various parameter sets were evaluated for the scattering pipeline. For $F_o$, 8 Hz, 16 Hz, 32 Hz and 64 Hz were tested. For these experiments, $Q_a=Q_e=8$. Additional experiments were conducted, in which all configurations for $Q_a \in \{1,2,4,8\}$ and $Q_e \in \{1,2,4,8\}$ were tested. 
\begin{figure}
\centering
\includegraphics[scale=\scalex]{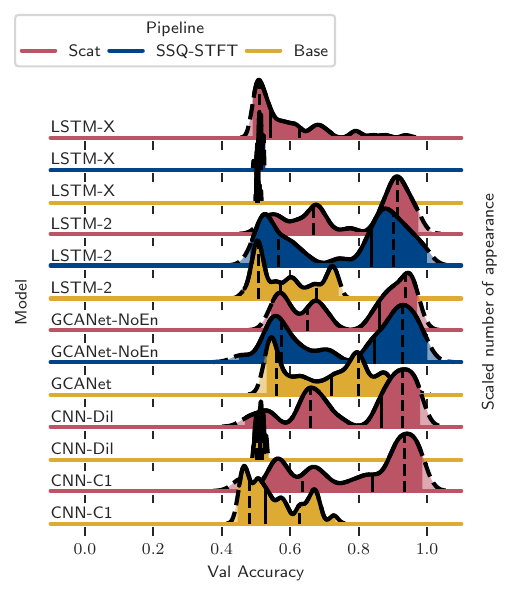} 
\caption{Kernel density plots comparing baseline (base), ST (scat), and SSQ-STFT preprocessing for speaker-wise evaluation on the KUL dataset. Accuracy distributions for trial-wise \CV{} across all subjects with $L_x=\SI{2}{\second}$
(CNN-C1, CNN-Dil) and $L_x=\SI{1}{\second}$ (others). ST parameters:$F_o=\SI{8}{\Hz}$, $Q_a=Q_e=8$.
}
\label{fig:ModelKULSpeakerwise}
\end{figure}
\subsubsection{Differences between the models}

In speaker-wise evaluation, all models benefit greatly from the ST, see \autoref{fig:ModelKULSpeakerwise}. The results of CNN-Dil, CNN-C1, GCANet-NoEn and LSTM-2 are close to each other, with LSTM-2 and GCANet-NoEn often performing slightly, but not significantly better according to the t-test. LSTM-X performs significantly worse than for the trail-wise evaluation. A comparison of \autoref{fig:ModelKULSpeakerwise} and \autoref{fig:ModelsKULTrialwise} also shows that CNN-Dil and CNN-C1 perform better when evaluated speaker-wise on the KUL dataset. This could be caused by overfitting when evaluating trial-wise. The speaker-wise evaluation only includes one speaker pair in training with the same number of attentions. 

\subsubsection{Influence of the second scattering layer}
\begin{figure}
\centering
\includegraphics[scale=\scalex]{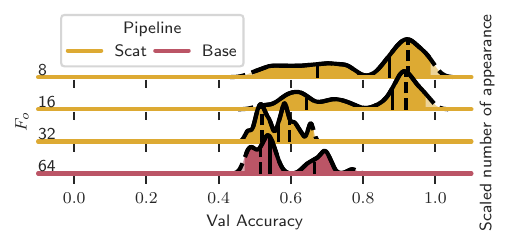} 
\caption{Collection of kernel density plots to compare the results of the scattering pipeline for different $F_o$ and the baseline pipeline for the speaker-wise evaluation on the KUL dataset, with $L_x=\SI{2}{\second}$ for the LSMT-2. Each plot shows the accuracy distribution of the different runs over all subjects. $Q_a=Q_e=8$ for the scattering pipeline.}
\label{fig:speaker_fs}
\end{figure}

The ST was tested with different values for $F_o$.
The second layer of the ST is only applied in the given implementation when $J>2$, which is only the case for the EEG data with $F_o<\SI{32}{\Hz}$, see Tables \ref{Tab:qfilter_audo} and \ref{Tab:qfilter_eeg}. 
\autoref{fig:speaker_fs} exemplary shows the results for the LSTM-2 with $L_x=\SI{2}{\second}$ and the baseline, where the same pattern can be observed for the other models. As one can see, for $F_o$ of 8 Hz and 16 Hz, a clear improvement is achieved when using the scattering pipeline, which is significant. For $F_o$ of 64 Hz and 32 Hz, however, the t-test showed no significant improvement. 

The LSTM-2 model achieved the best overall results of 0.88 for $F_o$ of 16 Hz and 8 Hz. For 16 Hz and 32 Hz, with $Q_e=Q_a=8$ and $L_x=\SI{2}{\second}$, the LSTM-2 achieved a median accuracy of 0.88 and 0.57, respectively. For $F_o=\SI{8}{\Hz}$, there is no significant difference. The question arises whether the improved performance is purely due to the number of parameters, or whether the second layer adds essential information.

Three observations indicate the latter: First, there is no increase in performance for $F_o=\SI{32}{\Hz}$ compared to the baseline while multiplying the number of channels by a factor of 19.
Second, high overfitting was observed when training with $F_o=\SI{32}{\Hz}$. While accuracies during training were further increasing to above 90\%, the performance during validation did not increase. This could indicate that the number of trainable parameters was already too high. 
Third, the $Q$ values were varied, and here, configurations having nearly the same number of parameters as with $F_o=\SI{32}{\Hz}$ but using an active second layer achieved much better results. These results will be described in detail in the next section.  
Overall, it can be seen that the second layer of the ST provides a significant additional information gain compared to a pure filter bank representing the ST without a second layer. In addition, it clearly outperforms traditional preprocessing methods. 

\subsubsection{Influence of the number of filters per octave}

\begin{table}
\centering
\caption{Comparison of the tested values for the scattering parameters $Q_a$ and $Q_e$ for the speaker-wise evaluation on the KUL dataset. The heatmap shows the median accuracies for the different configurations $Q_a$ and $Q_e$ for each model. $L_x=\SI{2}{\second}$ and $F_o=\SI{8}{\Hz}$.}
\includegraphics[scale=\scalexx]{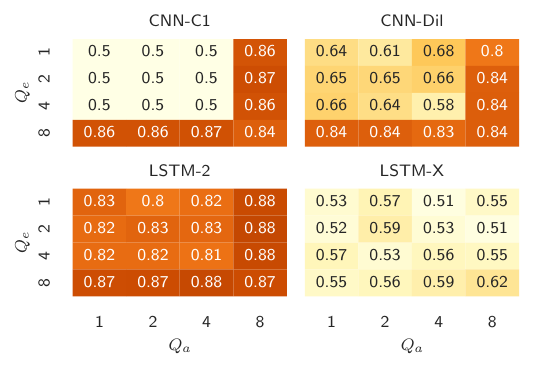} 
\label{fig:speaker_qs}
\end{table}
\begin{figure}
\centering
\includegraphics[scale=\scalexx]{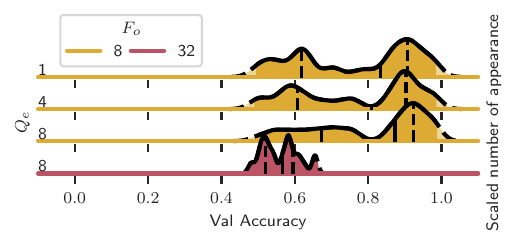} 
\caption{Comparison of the tested values for the scattering parameters $F_o$, $Q_a$, and $Q_e$ for the speaker-wise evaluation on the KUL dataset. $L_x=\SI{2}{\second}$.}
\label{fig:parameter_qs}
\end{figure}
In \autoref{fig:speaker_qs}, it can be seen that the choice of $Q_a$ and $Q_e$ has a measurable influence on the results. The number of resulting channels can be found in \autoref{Tab:qfilter_audo} and \autoref{Tab:qfilter_eeg}. \autoref{fig:speaker_qs} shows that reducing only one of the $Q$ values does not have a significant impact. When both values are reduced simultaneously, performance drops significantly for all models except LSTM-X. At up to 0.34, the decline is very high for CNN-Dil and CNN-C1, while it is not as high for LSTM-2, but still remains significant according to the t-test. After the initial decline when both values are reduced to four, performance remains constant, and further reduction appears to have no effect. 

Here, it can be shown once again that the observed performance does not depend directly on the number of parameters. In the previous experiment with $F_o=\SI{32}{\Hz}$ and $Q_e=Q_a=8$, there were a total of 1260 input channels, while in the current experiment with $F_o=\SI{8}{\Hz}$ and $Q_e=Q_a=1$, there are 1010, i.e., 250 fewer. Nevertheless, the LSTM-2 achieves a median of 0.82 compared to 0.57 with the smaller number of parameters, see \autoref{fig:parameter_qs}.

\subsubsection{Difference between subjects and runs}
\begin{figure}
\centering
\includegraphics[scale=\scalexx]{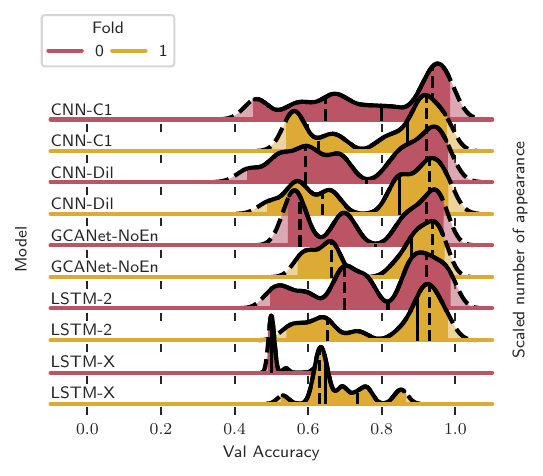}
\caption{Comparison of the results for the different speaker sets of the speaker-wise evaluation on the KUL dataset for different window lengths. Each of the kernel density plots shows the accuracies over all subjects for one of the speaker sets. The scattering pipeline with the LSTM-2 and $F_o=\SI{8}{\Hz}$, $Q_a=8$, and $Q_e=8$ was applied.}
\label{fig:speaker_split}
\end{figure}

The difference between the subjects is very high. There is one group of subjects where the achieved accuracies are between 0.4 and 0.6. For these subjects the models predict nearly random. There is a second group of subjects where the accuracies are between 0.75 and 1 on which the models predict mostly correct, see \autoref{fig:speaker_split}. Intermediate results are rarely seen, which indicates that the method is either working good for a subject or not at all. This raises the question of whether this behavior is due to the subject himself or the recording quality. There is, of course, further potential for improvement. 

\begin{figure}
\centering
\includegraphics[scale=1]{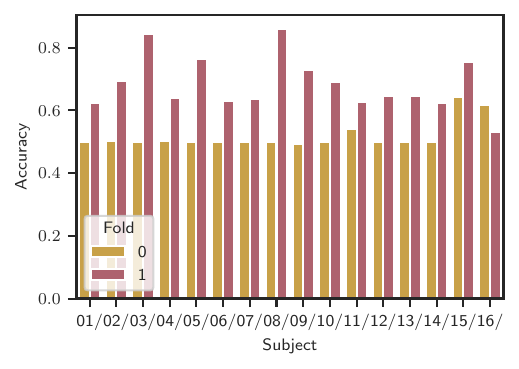} 
\caption{Comparison of the two different splits for the speaker-wise evaluation on the KUL dataset for each subject, for $L_x=\SI{2}{\second}$. Each line represents the accuracy for one run on one subject with one of the two speaker sets. The scattering pipeline with the LSTM-2 and $F_o=\SI{8}{\Hz}$, $Q_a=2$, and $Q_e=8$ was applied.}
\label{fig:speaker_split_bar}
\end{figure}

The results for the two different runs are compared in \autoref{fig:speaker_split}. In almost all experiments, the median values of the first run are worse than those of the second run. The difference in the mean values is much smaller. The difference between the two runs cannot be considered significant for all models, except for the LSTM-X model. Here, the difference is highly significant. In the first run, the results are not better than chance, while in the second run they are significant better than chance. An example can be found in \autoref{fig:speaker_split_bar}.

\subsection{Cross-Subject Evaluation}
For the cross-subject tests, the models were evaluated using the \CV{} method, in which the subjects are divided so that half of the subjects are used for training and the other half for evaluation. The preprocessing of the baseline is compared with the ST. For the scattering pipeline, only one parameter set with $F_o=\SI{8}{\Hz}$, $Q_a=8$, and $Q_e=8$ was tested. This configuration was chosen based on the results of the subject-based tests. As in the previous experiments, four different window lengths were evaluated. 

The cross-subject and speaker-wise evaluation can be combined. The subjects were divided according to the \CV{} scheme. For each split, four instead of two runs were performed, whereby the trials of one speaker pair were excluded from the training set and those of the other from the validation set. As described in Section \ref{sec:speaker-wise}, this evaluation could only be done with the KUL dataset.

\begin{figure}
    \centering
    \includegraphics[scale=\scalexx]{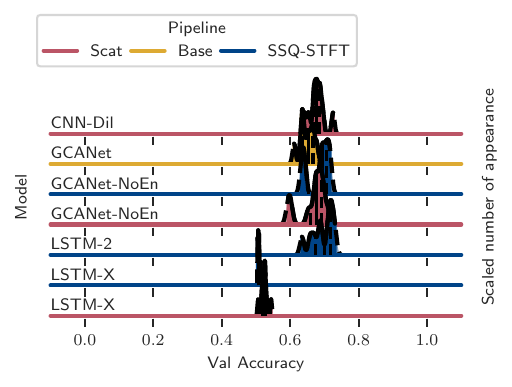} 
    \caption{Overview of the experimental results for the cross-subject tests on the KUL dataset for $L_x=\SI{1}{\second}$. Each kernel density plot shows the accuracy distribution over all subjects and runs of the \CV{}. The results for the models with the preprocessing piplines are compared. The scattering pipeline was applied with $F_o=\SI{8}{\Hz}$, $Q_a=8$ and $Q_e=8$.}
    \label{FIG:IntersubKULSpeaker}
\end{figure}

For the cross-subject tasks on the KUL dataset, no median accuracies greater than 0.7 could be achieved, see \autoref{FIG:IntersubKULSpeaker}. The patterns are identical to those of the subject-wise evaluation. A median accuracy of 0.7 for two-second windows is not as impressive as the 0.95 for the subject-wise evaluation, but it is still better than the CCA-based methods for the same and shorter window lengths \cite{geirnaert_electroencephalography-based_2021}. 
With a mean accuracy of 0.75, GCANet achieved slightly better results for the KUL dataset in the original publication with a leave-one-out cv (LOO), but no meaningful results for the DTU datasets, such as our models \cite{dai_gcanet_2026}.

\subsection{Models}
The ST exhibits varying compatibility across tested models. CNN-C1, an older architecture already surpassed by recent developments, performs poorly with ST due to its incompatibility with the reduced temporal representation a limitation shared by CNNs in general for this task. The dilation network faces similar issues, as its reliance on audio-EEG correlation becomes ineffective with dilation filters operating at 8 Hz output frequency.
In contrast, LSTM-based models and particularly GCANet-NoEn demonstrate strong ST compatibility. Replacing GCANet's encoding with ST substantially reduced trainable parameters, contributing to improved overall performance. Despite declining usage in current AAD research, LSTMs remain promising when combined with ST, especially given their 10-fold reduction in FLOPs compared to GCANet-NoEn.
\subsection{Evaluation Strategy}
\CV{} was employed as the primary evaluation method, allocating only half the data for training. This approach has important implications. As shown in \autoref{Tab:resultseavaluation}, training data quantity substantially affects performance, and limited training data may underestimate a model's true potential. This challenge is particularly acute in AAD, where per-subject data is inherently scarce.
However, this apparent limitation can be viewed as a strength for practical applications. A model that achieves robust performance with minimal training data is highly valuable in real-world scenarios, where the goal is to minimize per-subject data collection burden. The 5-fold repetition in \CV{} provides strong statistical power by evaluating performance across five different data splits, rigorously assessing model robustness through repeated statistical validation. LOO is commonly used in AAD literature, \CV{} is considered more suitable for evaluating AAD systems intended for deployment, as it better assesses practical viability under realistic data constraints while maintaining superior statistical reliability.

\begin{table}
    \centering
    \caption{Best results for the first test series of the subject-wise evaluation. It includes the maximum mean validation accuracies compared to the original published results}
    \begin{tabular}{c|c|c|c}
    Dataset &  Model & Mean \CV{} &  Mean 5-fold cv\\
    \hline
    DTU  & GCANet & 0.55 & 0.83 \cite{dai_gcanet_2026} \\
    DTU & GCANet-NoEn & 0.78 & 0.94 \\
    DTU & LSTM-2 & 0.6 & 0.84 \\
    \hline
    KUL  & GCANet & 0.73 & 0.92 \cite{dai_gcanet_2026} \\
    KUL & GCANet-NoEn & 0.91 & 92 \\
    KUL & LSTM-2 & 0.86 & 0.92 \\
    \end{tabular}
    \label{Tab:resultseavaluation}
    \end{table}
\subsection{Computitional Costs and Limitations}
\begin{table} [h]
\caption{Floating point operations and weights for 1-second inputs across different preprocessing methods. Baseline: one EEG signal ($64\times 128$) and two audio signals ($1\times 128$). GCANet: two EEG signals ($64\times 128$ and $64\times 10$) and two audio signals ($1\times 16000$. ST configurations: $F_o \in \{8, 16\}$
Hz with $Q=8$ for both audio and EEG.
 }
\centering

\begin{tabular}{|l|l|c|c|}
\hline
\textbf{Model} & \textbf{Variant} & \textbf{FLOPs} & \textbf{Weights} \\
\hline
GCANet      & Baseline      & 70M   & 3M     \\
\hline
GCANet-NoEn & Scat(8,8)     & 50M     & 1.3M    \\
GCANet-NoEn & Scat(16,16)   & 50M     & 1.4M    \\
GCANet-NoEn & SSQ-STFT      & 61M     & 1.3M    \\
\hline
LSTM-X      & Baseline      & 1.8M   & 7.0K    \\
LSTM-X      & Scat(8,8)     & 4M    & 237K    \\
LSTM-X      & Scat(16,16)   & 4.5M    & 125K    \\
LSTM-X      & SSQ-STFT      & 12M    & 330K    \\
\hline
LSTM-2      & Baseline      & 1.8M   & 7.7K    \\
LSTM-2      & Scat(8,8)     & 4M    & 237K    \\
LSTM-2      & Scat(16,16)   & 4.5M    & 125K    \\
LSTM-2      & SSQ-STFT      & 12M    & 331K    \\
\hline
CNN-Dil     & Baseline      & 1.5M    & 4.4K    \\
CNN-Dil     & Scat(8,8)     & 1.0M    & 54.9K   \\
CNN-Dil     & Scat(16,16)   & 1.6M    & 37K     \\
\hline
CNN-C1      & Baseline      & 1.5M    & 7.2K    \\
CNN-C1      & Scat(8,8)     & 1.7M    & 193K    \\
CNN-C1      & Scat(16,16)   & 2.0M    & 108K    \\
\hline
\end{tabular}
\label{tab:flops}
\end{table}
The ST requires approximately 44 million (M) FLOPs, 20M per audio and 4M for all 64 EEG channels for processing a 1-second window with standard settings ($Q=8$, $F_o=8$), see Tables \ref{Tab:qfilter_audo}, \ref{Tab:qfilter_eeg}, and \ref{tab:flops}. For EEG signals, this represents 4M additional FLOPs that must be added to conventional preprocessing steps. For audio signals, the ST replaces the envelope calculation, which requires approximately 5.5M FLOPs when implemented using a 28-band gammatone filterbank.

In GCANet, ST-based encoding saves 31 M FLOPs (20M from network encoding, 11M from envelope elimination), but requires 44M FLOPs, yielding 94M total versus 81M originally. The LSTM-2 with ST requires $\approx$ 50M FLOPs half of GCANet with comparable performance.  In contrast, the original LSTM-X with conventional preprocessing requires only $\approx$ 13M FLOPs but achieves significantly lower decoding accuracy.

However, several factors mitigate these costs: ST's most expensive operations (FFT/IFFT) are highly optimized on specialized hardware; sparse bandpass coefficients enable FLOP reduction without quality loss; channel reduction by fewer filters per octave $Q$ or lower $F_o$ decreases complexity for both the network and the ST.
Finally, hardware development is advancing rapidly, making it seem realistic in the future to run large networks such as GCANet. Commercial solutions such as the Phonak DEEPSONIC™ chip already achieve 7.7 GFLOPs/s according to the manufacturer \cite{phonak_ag_revolutionary_2024}. Current research on programmable AI accelerators such as the GAP9 system demonstrates the feasibility of complex real-time DNNs \cite{ismail_artificial_2025}. 
 
In addition to the required processing power, wavelet filters introduce processing delays: with $Q=8$ and
$F_o=8$ Hz, the longest EEG path delays by 64 samples (0.5s). Including network computation, total delay may reach 1s. This can be reduced by increasing $F_o$ to 16 Hz or omitting low frequencies. Audio signals experience comparable delays, but these are less critical since the EEG response inherently lags behind the auditory stimulus. 

Additional limitations persist: ST does not improve cross-subject generalization—a known AAD challenge \cite{dai_gcanet_2026}. CNN-C1, LSTM-2, GCANet, and GCANet-NoEn currently support only two audio signals; LSTM-X addresses this but underperforms on KUL data. Adaptations for practical deployment are necessary and feasible.  

\section{Conclusion}
In this work, the application of the ST on the AAD task was proposed and validated. Comparisons were made to the actually common preprocessing pipelines. For this, a number of experiments have been conducted, considering different configurations for the ST and models for different AAD scenarios.
Subject-wise as well as cross-subject classification were included, both done on the two widely used datasets from DTU and KUL. Additionally, the scenario of unknown speakers was tested on the KUL dataset. Multiple t-tests were carried out in order to be able to reliably compare the results for different configurations and models. 

It was shown that the performance highly depends on the datasets and evaluation methods. Only the GCANet-NoEn in combination with the ST showed good performance on both datasets and all evaluation strategies.
Especially on the DTU dataset, most methods do not work well. This underlines that machine learning approaches are not good at generalization for AAD and that robust evaluation strategies such as trial-wise \CV{} should be performed in the future to not overestimate the performance. 

Across all subject-wise experiments, ST consistently delivered the best results. However, ST did not lead to neural-network functioning well across subjects. Nevertheless, it was shown that multi-layer ST can bring about a significant improvement compared to baseline preprocessing, regular filter banks, and SSQ-STFT.

\printbibliography
\end{document}